# HIGH RESOLUTION BPM UPGRADE FOR THE ATF DAMPING RING AT KEK*

N. Eddy#, C. Briegel, B. Fellenz, E. Gianfelice-Wendt, P. Prieto, R. Rechenmacher,
A. Semenov, D. Voy, M. Wendt, D. Zhang, Fermilab, Batavia, IL 60510, U.S.A.
N. Terunuma, J. Urakawa, KEK, Tsukuba, Japan

*Abstract*

A beam position monitor (BPM) upgrade at the KEK Accelerator Test Facility (ATF) damping ring has been accomplished, carried out by a KEK/FNAL/SLAC collaboration under the umbrella of the global ILC R&D effort. The upgrade consists of a high resolution, high reproducibility read-out system, based on analog and digital down-conversion techniques, digital signal processing, and also implements a new automatic gain error correction schema. The technical concept and realization as well as results of beam studies are presented.

## INTRODUCTION

The next generation of linear colliders require ultra-low vertical emittance of <2 pm-rad. The damping ring at the KEK Accelerator Test Facility (ATF) is designed to demonstrate this mission critical goal [1][2]. A high resolution beam position monitor (BPM) system for the damping ring is one of the key tools for realizing this goal.

The BPM system needs to provide two distinct measurements. First, a very high resolution (~100-200nm) closed-orbit measurement which is averaged over many turns and realized with narrowband filter techniques - "narrowband mode". This is needed to monitor and steer the beam along an optimum orbit and to facilitate beam-based alignment to minimize non-linear field effects. Second, is the ability to make turn by turn (TBT) measurements to support optics studies and corrections necessary to achieve the design performance. As the TBT measurement necessitates a wider bandwidth, it is often referred to as "wideband mode".

The BPM upgrade was initiated as a KEK/SLAC/FNAL collaboration in the frame of the Global Design Initiative of the International Linear Collider. The project was realized and completed using Japan-US funds with Fermilab as the core partner.

## THE ATF DAMPING RING

The Accelerator Test Facility at KEK consists of an S-Band electron linac, the damping ring and an extraction beam-line (ATF2). The machine parameters for the damping ring are shown in Table 1. There are 96 button-style BPM pickups in the The 1.2 GeV ATF damping ring is equipped with 96 button-style BPM pickups.

___________________________________________
* Work is supported by the joint high energy physics research program of Japan-USA, and by FNAL, operated by Fermi Research Alliance LLC under contract #DE-AC02-07CH11359 with the US D.O.E.
#eddy@fnal.gov

Table 1: ATF DR Machine Parameters

| | | |
|---|---|---|
| beam energy E | = | 1.28 GeV |
| beam intensity, single bunch | ≈ | ~1.6 nC ≡ $10^{10}$ e⁻ ($\equiv I_{bunch} \approx 3.46$ mA) |
| beam intensity, multibunch (20) | ≈ | ~22.4 nC ≡ 20 x 0.7 $10^{10}$ e⁻ ($\equiv I_{beam} \approx 48.5$ mA) |
| $f_{RF}$ | = | 714 MHz ($\equiv t_{RF} \approx 1.4$ ns) |
| $f_{rev}$ | = | $f_{RF}/330 \approx 2.16$ MHz ($\equiv t_{rev} \approx 462$ ns) |
| bunch spacing $t_{bunch}$ | = | $2/f_{RF} \approx 2.8$ ns |
| batch spacing | = | $t_{rev}/3 = 154$ ns |
| repetition freq. $f_{rep}$ | = | 1.56 Hz ($\equiv t_{rep} = 640$ ms) |
| beam time $t_{beam}$ | = | 460.41 ms ($\equiv$ 996170 turns) |
| vert. damping time $\tau$ | ≈ | 30 ms |
| hor. betaron tune (typ.) | ≈ | 15.204 ($\equiv f_h \approx 441$ kHz) |
| vert. betaron tune (typ.) | ≈ | 8.462 ($\equiv f_v \approx 1$ MHz) |
| synchrotron tune | ≈ | 0.0045 ($\equiv f_s = 9.7$ kHz) |

During standard operation a single bunch is injected into the damping ring. After approximately 200ms the beam is fully damped. For 1.56Hz operation beam remains in the damping ring for up to 460ms before being extracted. In multi-batch operation up to three batches containing from 1 to 20 bunches can be injected. The bunch spacing within a batch is 2.8ns. The BPM system provides closed-orbit measurements for all bunch configurations but the TBT measurements are only supported in the single bunch configuration.

## THE ATF DAMPING RING BPM UPGRADE

The initial electronics and read-out system for the damping ring BPMs did not meet the requirements in terms of resolution, reproducibility, or intensity independence nor did it provide turn by turn measurement capabilities which hindered the performance of the damping ring. In 2006 an effort was initiated to uprade the electronics based upon analog downmix technology developed at SLAC and digital signal processing via Echotek digitzers used in the Fermilab Tevatron BPM system. The system was prototyped using 20 of the 96 bpms over several years. The original SLAC design for the downmix electronics was modified to include a calibration and control module and the commercial Echotek digital receiver was replaced with a custom digitizer developed at Fermilab. An overview of the final system upgrade is shown in Figure 1.

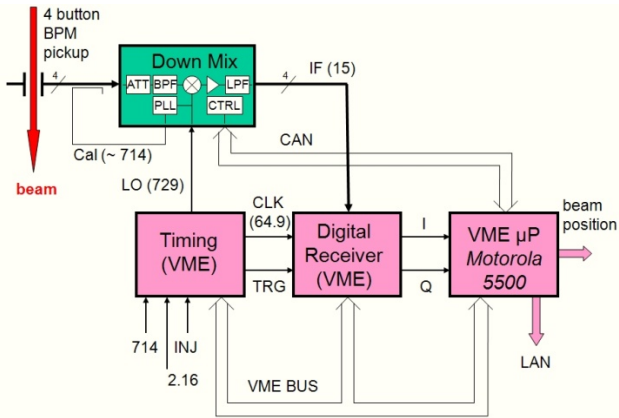

Figure 1: ATF DR BPM System Overview.

## Hardware

The downmix electronics are located in the tunnel right next to each BPM pickup. The upgraded downmix box now includes programmable control of the gain and attenuation stages as well as complete control of the generation of the calibration tone via a frequency synthesizer. The calibration tone is only on during the narrowband measurement. It is turned off at injection and extraction to facilitate turn by turn measurements. The custom timing module receives the machine 714MHz RF and generates the LO for the downmix boxes as well as the clock for the digitizers. Details of both modules can be found in [3,4].

The Echotek digital receiver has been replaced with a custom digitizer developed at Fermilab which is designed specifically for accelerator applications and allows for fully customizable firmware specific to the given task. A block diagram of the digitizer is shown in Figure 2.

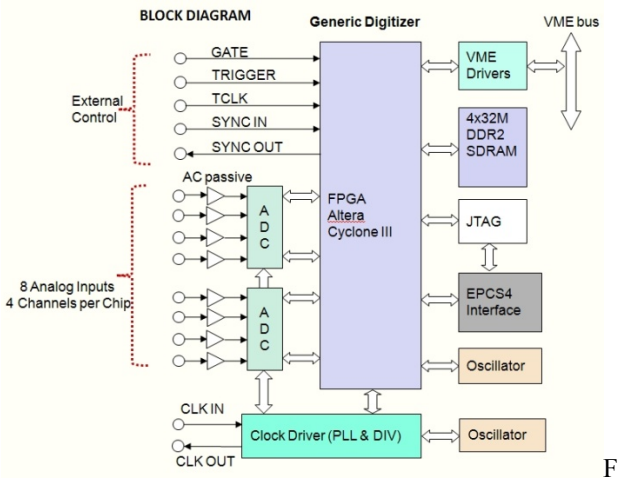

Figure 2: Block diagram of the custom digitizer.

For the ATF BPM upgrade, the custom board allows us to perform all the signal processing for the closed-orbit and turn by turn measurements in parallel as well as provide provide diagnostic data such as I,Q pairs and raw ADC data for every trigger. A simplified block diagram of the digitizer firmware for the ATF BPM is shown in Figure 3.

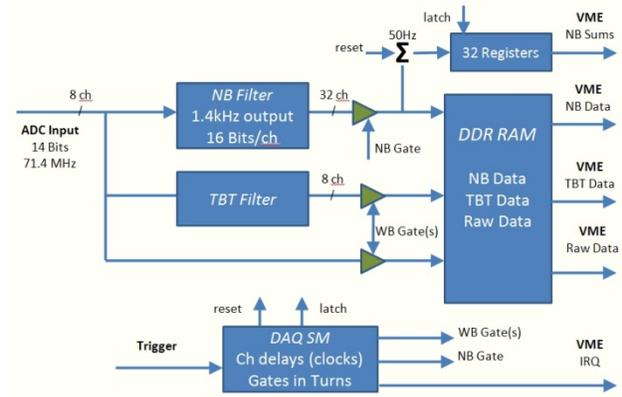

Figure 3: Block diagram of the ATF digitizer firmware.

The closed-orbit measurement is realized by a narrowband filter based upon quadrature digital downconversion. It is implemented in the firmware as shown in Figure 4. The key components are a five stage cascade integrating comb (CIC) filter and an 80 tap finite impulse response (FIR) filter which process inphase (I) and quadrature (Q) components of the signal. The CIC filter processing I,Q for the beam and calibration signals in parallel while the FIR filter makes use of the high clock rate to process all 32 CIC outputs in series to conserve resources. Diagnostic peak detectors are implanted to monitor for digital saturation. The magnitudes for beam and calibration signals are calculated in the firmware.

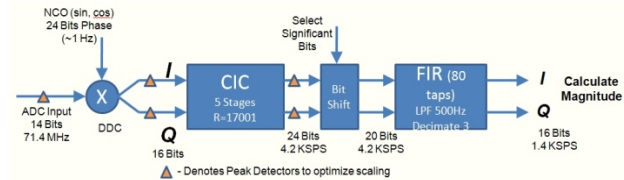

Figure 4: Narrowband signal processing.

With the digitizers operating at 71.4MHz, we get exactly 33 samples per turn. The turn by turn filter is simple box integrator over a programmable subset of the 33 samples. The algorithm is equivalent to summing the I,Q magnitudes over the specified interval. The number of samples is chose to optimize the signal to noise.

## Software

A simplified block diagram for the control and readout software is shown in Figure 5. All interaction with the front-end is handled via the EPICS IOC. Under normal operation, all relevant physics data along with some select diagnostic data are read into the Motorola 5500 cpu. For standard physics mode, the following data is made available for all bpms:

- Injection First Turn (position & intensity)
- Injection Turn by Turn for 1024 turns (position & intensity)
- Closed Orbit (position & intensity along with rms)
- Extraction Last Turn (position & intensity)

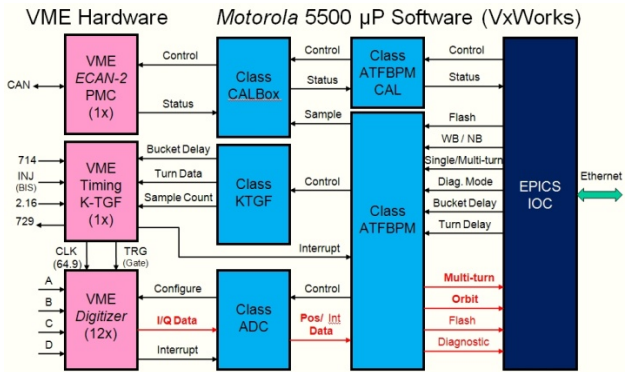

Figure 5: Software overview.

For dedicated turn by turn studies, the beam is pinged after being damped. In this special case, the calibration tone and closed orbit measurement are disabled and only the TBT measurement at the time of the excitation is made.

The front-end readout and calculation of positions and intensities takes < 20ms. For TBT and diagnostic data, it is not possible to return all the data over the network at 1.56Hz. For these cases, a synchronized snapshot is implemented which stores all data in the front-end for the same injection trigger and can be readout at leisure. The snapshot is used to collect TBT data from all BPMs while the system continues making measurements.

## PRELIMINARY RESULTS

The calibration system was checked by inserting a 1 dB attenuator on a single channel for a select bpm. With no calibration, this results in a 265mm position change both horizontal and vertical. With the calibration, there is no change in the positions as expected.

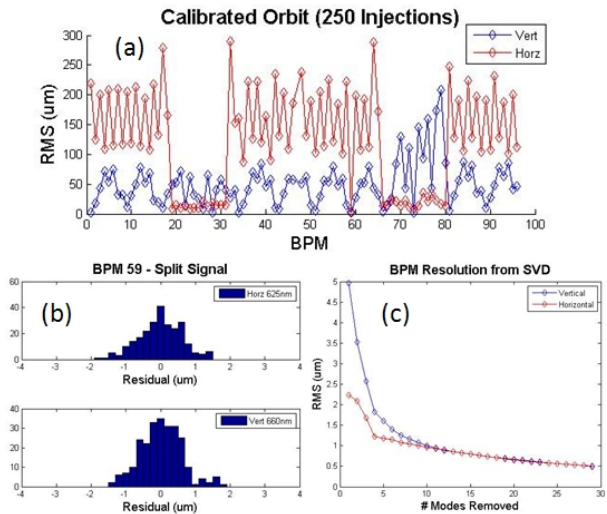

Figure 6: (a) Raw BPM orbit RMS over 8 hours. (b) Split BPM resolutions. (c) SVD resolutions.

The closed-orbit resolution was estimated in two ways. First, a single button electrode from BPM 59 was put to a four-way splitter thus all four channels for this BPM received a "real" beam signal which had intensity variations but should have a fixed position. Over an 8 hour shift, 250 orbit measurements were made. The RMS on the position at each BPM is shown in Figure 6. The resolution can also be estimated using the Singular Value Decomposition (SVD) technique [5]. The results for both are shown in Figure 7. The split signal results in about a 650nm resolution. Ideally, we expect to get the same results from the SVD technique after removing correlated beam motion – expected to be 4-6 modes. In principle, the resolution of the bpms should be the same for horizontal and vertical, but the first 10 SVD modes show substantial differences. Further, it is necessary to remove 20 SVD modes to achieve the 650nm resolution. The SVD results are still under study. It was also found after these studies that the downmix boxes gain was -10dB from the nominal setting. One would expect a factor ~3 improvement on the resolution for the proper gain setting.

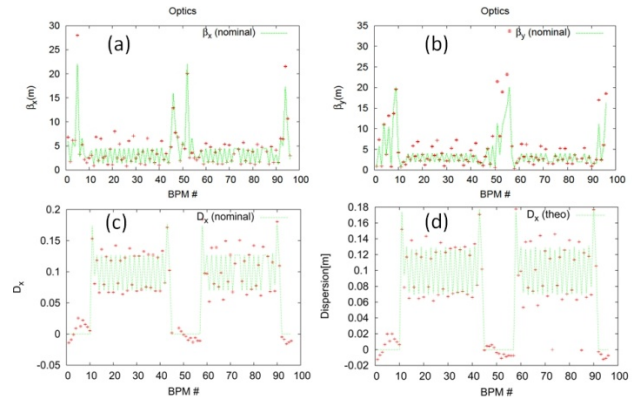

Figure 7: (a) Horizontal β function. (b) Vertical β function. (c) Horizontal dispersion from TBT at injection. (d) Horizontal dispersion from closed-orbit.

A number of studies were also carried out to examine the ring optics. For the TBT studies a pinger is fired in both planes after the beam is well damped. The resulting beam response for the next 1024 turns of data are collected from all bpms. The dispersion was measured from both injection TBT data synchrotron oscillations and from the orbit change due to changing the ring RF. The results for each measurement as shown in Figure 8 are in reasonable agreement with the design optics.

## SUMMARY

A complete upgrade of the ATF damping ring BPM electronics and readout has been completed. The system is now fully operational. It's performance will be further studied once the ATF facility resumes operation.